\documentclass[aps,preprint,showpacs,showkeys]{revtex4}
\usepackage{graphicx}
\def\p{\textrm{\bf p}}
\def\q{\textrm{\bf q}}
\mathchardef\G="3247
\mathchardef\T="3254

\begin{document}

\title{Microscopic analog of temperature within nonextensive thermostatistics}
\author{M. P. Almeida, F. Q. Potiguar, U. M. S. Costa}

\affiliation{Departamento de F\'{\i}sica, Universidade Federal do Cear\'a,
60455-900 Fortaleza, Cear\'a, Brazil}

\begin{abstract}
It is presented a microscopic interpretation for the
temperature within Tsallis thermostatistics,  generalizing the classical
derivation based on the Boltzmann-Gibbs statistics.
It is shown that with this definition the zeroth law and the equipartition theorem are valid in their classical form.
Moreover, it is observed that the equation of state for an ideal gas within generalized thermostatistics
preserves the classical Boyle's law form $PV=NkT$.
\end{abstract}

\maketitle

\section{Introduction}

The explanation of the macroscopic theory of thermodynamics in terms of the
more abstract microscopic statistical mechanics is one of the major
achievements of the physics of the early twentieth century.
This task included
the establishment of connections between the macroscopic {\it state variables} and
their microscopic counterparts.
Among the thermodynamics variables, the {\em external
parameters},
e.g. the volume of a gas, and the mechanical related
quantities, e.g. the pressure, are easily interpreted microscopically.
However, the  non-mechanical thermodynamical
quantities, viz., temperature, entropy and their various subsidiary quantities,
are more cumbersome to identify microscopically, since they are basically
defined within  thermodynamics, with no apparent macroscopic mechanical
equivalent quantity.
The statistical mechanical analogues of temperature and entropy
are characteristic quantities of the ensemble distribution.

The concept of temperature of a body lies in the very root upon which
the whole science of thermodynamics rests.
It is connected to our common sense
of ``hot'' and ``cold'', and the zero law of thermodynamics is the statement
that this quantity can be compared by observing the behavior of bodies put
in contact to each other.
In statistical mechanics the temperature can be related, via the theorem of equipartition of energy,
to the amount of kinetic energy per degrees of freedom of the system.
Within the Boltzmann-Gibbs (BG)  approach, it is also associated with the $\beta$ constant that appears
in the exponent of the canonical distribution.

An alternative formalism to the Boltzmann-Gibbs thermodynamics development
has been evolved in the last decade starting from the postulate of a
generalized entropy form by Tsallis \cite{tsallis88}.
Since then, a wealthy scientific bibliography has been produced presenting theoretical and
experimental confirmation of the generalized thermostatistics.
For a full and updated list of publications check \cite{www}.

It  has been pointed out by Plastino \cite{PlasPlas94} that systems with a
small number of particles obey the generalized thermostatistics.
Moreover, it has been proven by Almeida \cite{Alm01} that both formalisms
(Boltzmann-Gibbs and Tsallis) can be obtained via a common logical development
based on the hypothesis of equiprobability on phase space, ergodicity and the
use of structure functions.

Our goal in this paper is to present a microscopic interpretation for the
temperature within Tsallis thermostatistics,  generalizing the classical
interpretation derived on the Boltzmann-Gibbs formulation.
This endeavor has been already considered by Martinez et. al. \cite{PhysA295_246, PhysA295_416}
and Plastino et. al. \cite{PLA260_46}. However, our presentation, which is based on elementary
techniques os statistical mechanics, leads to the resolution of disparities with the classical
Boltzmann-Gibbs approach that were encountered in these papers.

\section{Preliminary considerations}
Consider a system $\G$ with Hamiltonian $H=H(\q,\p)$ and
let $\Xi$ denote its phase space. Let $V(E)$ be the volume of the region of the phase-space where
$H(\q,\p)<E$, and let $\Omega(E)= \partial V(E)/\partial E$ denote the
{\em structure function} \cite{Kinchin}, also known as density of states.
Let's define the function
\begin{equation}\label{beta}
\beta(E)=\frac{d}{dE}\ln\Omega(E).
\end{equation}

Let the system $\G$ be composed by two subsystems $\G_1$ and $\G_2$ with phase spaces $\Xi_1$ and
$\Xi_2$, respectively. For $i=1,2$, let's denote by $V_i(E)$ the volume function in the phase space
$\Xi_i$ of the region where $H_i(\q_i,\p_i)<E$,  $\Omega_i(E)$ its respective structure function and
$\beta_i(E)$ the associated beta function.

If $\G_1$ and $\G_2$ are in {\em weak interaction}
with each other, in the sense that $H(\q,\p)= H_1(\q_1,\p_1) + H_2(\q_2,\p_2)$, i.e., there is no
term of interaction of $\G_1$ with $\G_2$ in the Hamiltonian, then the canonical distribution of the system
$\G_1$ in its phase space $\Xi_1$, when the total energy of $\G$ is $E=a$, is given by the density
function \cite{Kinchin}
\begin{equation} \label{rho_1}
\rho_1(E_1)=\frac{\Omega_2(a-E_1)}{\Omega(a)},
\end{equation}
and the structure function of the system $\G$ satisfies the convolution relation
\begin{equation}\label{convolution}
\Omega(E)=\int_0^E \Omega_1(\xi)\Omega_2(E-\xi) d\xi.
\end{equation}

It has been shown in \cite{Alm01} that the form of $\rho_1$ depends on the value of the parameter $q_2$
determined by
\begin{equation} \label{dbeta}
\frac{d}{dE}\left(\frac{1}{\beta_2(E)}\right)=q_2-1.
\end{equation}
When $q_2=1$, $\rho_1$ is an exponential (Boltzmann-Gibbs) and, when $q_2\ne 1$, $\rho_1$ is a power-law
(Tsallis) distribution. The function $V_2(E)$ can be obtained integrating
Eq.~(\ref{dbeta}). Let's assume that all of the volume functions $V_i(E)$ are such that $V_i(0)=\Omega_i(0)=0$.
Therefore, the general form of $V_2(E)$ is
\begin{equation}\label{V2}
V_2(E)=C_2 E^{q_2/(q_2-1)},
\end{equation}
with $C_2$ a constant.
An analogous procedure for the subsystem $\G_1$ yields
\begin{equation}\label{V1}
V_1(E)=C_1 E^{q_1/(q_1-1)},
\end{equation}
with $C_1$ a constant.
Hence, for $i=1,2$,
\begin{equation}\label{Omega_i}
\Omega_i(E)= \frac{q_i C_i}{q_i-1} E^{1/(q_i-1)},
\end{equation}
and
\begin{equation}
\beta_i(E)=\frac{1}{(q_i-1)E}.
\end{equation}
From Eq.~(\ref{convolution}) we get that
\begin{equation}\label{Omega}
\Omega(E)= C_1 C_2 \frac{\Gamma\left(\frac{q_1}{q_1-1}+1\right)
 \Gamma\left(\frac{q_2}{q_2-1}+1\right)}{\Gamma\left(\frac{q}{q-1}\right)} E^{1/(q-1)},
\end{equation}
where
\begin{equation} \label{qrelation}
\frac{q}{q-1}=   \frac{q_1}{q_1-1}+\frac{q_2}{q_2-1},
\end{equation}
 and $\Gamma(x)$ is the gamma function. The function $\beta$ for $\G$ is then given by
\begin{equation}
\beta(E)=\frac{1}{(q-1)E}.
\end{equation}
From Eq.~(\ref{Omega}) we get that
\begin{equation}\label{V}
V(E)=  C_1 C_2 \frac{\Gamma\left(\frac{q_1}{q_1-1}+1\right)
 \Gamma\left(\frac{q_2}{q_2-1}+1\right)}{\Gamma\left(\frac{q}{q-1}+1\right)} E^{q/(q-1)}.
\end{equation}

The interpretation of the temperature is exactly the one based
upon the theorem of equipartition of energy for the system $\G$ with total energy $E=a$,
which might be expressed \cite{Huang} in the form
\begin{equation}\label{equipartition}
\left< x_i\frac{\partial H}{\partial x_j}\right> = \delta_{ij}\frac{V(a)}{\Omega(a)} = \delta_{ij} kT
\end{equation}
where $x_i$ stands for any of the variables $p_i$ or $q_i$, $k$ is the Boltzmann constant
and $T$ is the absolute temperature. This is the same temperature definition for the Boltzmann-Gibbs
setting.
Using the expressions for $V(a)$ and $\Omega(a)$, Eqs.~(\ref{V}) and (\ref{Omega}) respectively,
and the definition of $\beta$, Eq.~(\ref{beta}), we may write the temperature of
the system $\G$ as
\begin{equation}\label{kT}
T= \frac{1}{k}\left(\frac{q-1}{q}\right) a=\frac{1}{kq\beta(a)}.
\end{equation}

For a general system, we postulate that the function
\begin{equation}\label{Tdefinition}
\T(\G)=\frac{1}{kq}\left<\frac{1}{\beta(E)}\right>
\end{equation}
represents the physical temperature of the system $\G$.
Note that $\T$ is a function that depends on the observed part of the system.
For the whole system $\G$ with constant energy $E=a$, we have that $\beta(a)=((q-1)a)^{-1}$ and
\begin{equation}\label{TG}
\T(\G) = \frac{1}{k}\left(\frac{q-1}{q}\right) a =T .
\end{equation}
Consider now the subsystem $\G_1$, which has a random energy $E_1$ distributed according to
a probability density function \cite{Kinchin}[\S 15, pp. 75]
\begin{equation}
\frac{\Omega_1(x)\Omega_2(a-x)}{\Omega(a)}.
\end{equation}
Therefore, the average value of the energy $E_1$ is
\begin{eqnarray}\label{meanE_1}
\left<E_1\right>& = &\frac{1}{\Omega(a)}\int_0^a x \Omega_1(x) \Omega_2(a-x) dx \\
                & = & \frac{B\left(\frac{q_1}{q_1-1} +1, \frac{q_2}{q_2-2}\right)}
                { B\left(\frac{q_1}{q_1-1}, \frac{q_2}{q_2-2}\right)} a \\
                & = & \left(\frac{q_1}{q_1-1}\right)\left( \frac{q-1}{q}\right) a,
\end{eqnarray}
where $B(m,n)$ is the beta function.
Then, from Eq.~(\ref{Tdefinition}) we get the value
\begin{eqnarray}
\T(\G_1) & = & \frac{1}{kq_1}\left<\frac{1}{\beta(E_1)}\right> \\
         & = & \frac{1}{k} \left(\frac{q_1-1}{q_1}\right) \left<E_1\right> = \frac{1}{k}
         \left(\frac{q-1}{q}\right) a \\
         & = & \T(\G),
\end{eqnarray}
which shows that the evaluation of $\T$ over any of the subsystems $\G_i$ produces the same result.

Let's consider now the combination ({\em weakly interaction}) of two systems $\G_1$ and $\G_2$, with parameters $q_1$ and $q_2$
respectively, to form a compound system $\G$.
Let's impose the condition that before the assembling $\T(\G_1)=\T(\G_2)$, i.e.,
\begin{equation}\label{inicond}
\left(\frac{q_1-1}{q_1}\right)a_1= \left(\frac{q_2-1}{q_2}\right)a_2,
\end{equation}
where $a_1$ and $a_2$ are the initial energies of the systems $\G_1$ and $\G_2$ respectively.
Thus, the compound system $\G$ has energy $a=a_1+a_2$ and parameter $q$ related to $q_1$ and $q_2$
by Eq.~(\ref{qrelation}). Therefore, computing $\T(\G)$ using Eq.~(\ref{Tdefinition}) we have
\begin{eqnarray}
\T(\G) = \frac{1}{k}\frac{a}{\left(\frac{q}{q-1}\right)}  = \frac{1}{k}\left(\frac{a_1+a_2}
{\frac{q_1}{q_1-1} + \frac{q_2}{q_2-1}}\right)& = & \frac{1}{k}\frac{a_1}{\left(\frac{q_1}{q_1-1}\right)}
=\T(\G_1) \\
& = & \frac{1}{k}\frac{a_2}{\left(\frac{q_2}{q_2-1}\right)} =\T(\G_2),
\end{eqnarray}
which implies that $\T(\G)$, considered as a temperature scale, satisfies the {\em zero law of thermodynamics}.

\section{Application}
Consider now a system composed by $N$ free particles in a confined three-dimensional volume $V$, i.e.,
an ideal gas system.
It has been stated by Plastino and Lima \cite{PLA260_46} that the virial theorem in the
generalized thermostatistics context leads to an equation of state for the ideal gas that is different
from the usual Boyle's law; The Boyle's law being recovered in the limit case of $q\to 1$.
This conflicting result was generated by the adoption of the temperature $T$ as inverse of
$\beta$ divided by the Boltzmann constant, $T=(k\beta)^{-1}$. If instead of this we define the temperature
by the relation Eq.~(\ref{Tdefinition}) we assure a priori the validity of the equipartition theorem
in its classical form Eq.~{\ref{equipartition}). The classical virial theorem for the
ideal gas \cite{Pathria} states that
\begin{equation}\label{gas_virial}
\sum_{i}q_i F_i = -P \oint_S {\bf r}\cdot d{\bf S}= -3PV,
\end{equation}
where ${\bf r}$ is the position vector of a particle and $F_i$ is the generalized $i$ force associated to
$q_i$.
Therefore, the definition of temperature by Eq.~(\ref{Tdefinition}) together with the classical virial theorem
leads to the Boyle's law in its original form,
\begin{equation}\label{Boyle}
PV=NkT,
\end{equation}
even in the context of the generalized thermostatistics.

\section{Conclusion}

We have presented a microscopic identification of the temperature valid for both
Boltzmann-Gibbs and Tsallis statistics. It was observed that the zeroth law of thermodynamics, the theorem of equipartition,
and the equation of state for ideal gas are all valid in their classical form even within Tsallis framework, provided the 
adequate definition of temperature is taken.
The adoption of an inadequate form of the temperature leads to conflicting results obtained by the BG and the Tsallis 
thermostatistics, e.g. the modified Boyle's law deduced in \cite{PLA260_46}.

\section{Acknowledgments}
F. Q. Potiguar was financially supported by FUNCAP; M. P. Almeida and U. M. S. Costa were partially supported by CNPq.


\begin{thebibliography}{99}

\bibitem{tsallis88} C. Tsallis, J. Stat. Phys. {\bf 52} (1988) 479 .

\bibitem{www} http://tsallis.cat.cbpf.br/biblio.htm

\bibitem{PlasPlas94}A. R. Plastino, A. Plastino,
Phys. Lett. A {\bf 193} (1994) 140.

\bibitem{Alm01}M. P. Almeida,
Physica A {\bf 300} (2001) 424.

\bibitem{PhysA295_246}
S. Martínez, F. Pennini, A. Plastino, Physica A, {\bf 295} (2001) 246.

\bibitem{PhysA295_416}
S. Martínez, F. Pennini, A. Plastino, Physica A, {\bf 295} (2001) 416.

\bibitem{PLA260_46}
A. R. Plastino, J. A. S. Lima, Phys. Lett. A {\bf 260} (1999) 46.

\bibitem{Kinchin} A. I. Kinchin, {\it Mathematical Foundations of Statistical
Mechanics} (Dover, New York, 1949).

\bibitem{Huang} K. Huang, {\it Statistical Mechanics}, (John Wiley \& Sons, Inc.,
New York, 1963).

\bibitem{Pathria} R. K. Pathria, {\it Statistical Mechanics}, First Edition,
(Pergamon Press, Oxford, 1972). 

\end{thebibliography}
\end{document}